\documentclass[%
 aps,
 jmp,%
 amsmath,amssymb,
preprint,%
]{revtex4-1}

\usepackage{graphicx}
\usepackage{dcolumn}
\usepackage{bm}
\usepackage{epstopdf}
\usepackage{amssymb}
\usepackage{color}
\usepackage[colorlinks=true, letterpaper=true, pdfstartview=FitV, linkcolor=blue, citecolor=blue, urlcolor=blue]{hyperref}
\usepackage{lineno}
\usepackage[caption=false]{subfig}%

\begin{document}

\title{Multifunctional Altermagnet with Large Out-of-Plane Piezoelectric Response in Janus V$_{2}$AsBrO Monolayer}
\author{Qiuyue Ma}
\author{Busheng Wang}
\author{Guochun Yang}
\author{Yong Liu}\email{yongliu@ysu.edu.cn}
\affiliation{State Key Laboratory of Metastable Materials Science and Technology \& Hebei Key Laboratory of Microstructural Material Physics, School of Science, Yanshan University, Qinhuangdao 066004, China}

\begin{abstract}
Altermagnetism has emerged as a third fundamental category of collinear magnetism, characterized by spin-splitting in symmetry-compensated collinear antiferromagnets, opening new frontiers in spintronics and condensed matter physics. Here, based on first-principles calculations, we propose a novel altermagnetic semiconductor, the asymmetric Janus V$_2$AsBrO monolayer, which exhibits a magnetic easy axis favoring the out-of-plane direction and a N\'{e}el temperature ($T_N$) exceeding room temperature. The system exhibits a strain-tunable piezovalley effect, generating valley polarization under uniaxial strain. Notably, hole doping under uniaxial strain generates a net magnetization ($M$) through a piezomagnetic mechanism. Additionally, the broken inversion symmetry endows the monolayer with a substantial out-of-plane piezoelectric coefficient $d_{31}$ (2.19 pm/V), presenting broad prospects for the development and design of novel piezoelectric devices. Our findings provide a promising candidate material for the advancement of 2D multifunctional devices in nanoelectronics, spintronics, valleytronics, and piezoelectrics.
\end{abstract}

\maketitle


\maketitle
\section{Introduction}

Two-dimensional (2D) antiferromagnetic (AFM) materials are highly promising candidates for next-generation devices and technologies due to their robustness against magnetic disturbances, ultrafast dynamics, absence of stray fields, higher speed operation, and lower-energy consumption~\cite{Baltz2018,Han2023,Wang2023}. However, the counterpart spin degeneracy in real and reciprocal space of conventional antiferromagnets hinders the realization of spin-polarized currents. Recently, a new type of collinear magnetism called `altermagnetism' has garnered significant attention in the fields of spintronics and magnetic materials research~\cite{Karube2022,Gonz2021,mejkal2020,Yuan2020}. Altermagnetic materials exhibit spin splitting without relying on relativistic spin-orbit coupling (SOC) by breaking $\mathcal{PT}$ symmetry, in which $\mathcal{P}$ is spatial inversion symmetry and $\mathcal{T}$ is time-reversal symmetry~\cite{Hayami2020,RHe2023,Zhu2024,Reimers2024}, and maintaining AFM order with no net magnetic moment. The novel physical phenomena observed in altermagnets, such as giant and tunneling magnetoresistance~\cite{Shao2021,lHellenes2022}, spin current generation~\cite{HBai2022,MNaka2019}, anomalous Hall effect~\cite{ZFeng2022}, and suppressed Kondo effect~\cite{GDiniz2024}, are prompting the development of multifunctional electronic devices.

Manipulating energy valleys in the conduction/valence bands to encode, transfer, and store information is characterized as valleytronics~\cite{JSchaibley2016,GPacchioni2020,SVitale2018,SSattar2022}. In recent years, various 2D materials, such as graphene and the 2$H$-phase transition metal dichalcogenides (TMDs) MX$_2$, have been identified as promising candidates for valleytronic applications. In these materials, the $K$ and $K$' valleys are related through $\mathcal{T}$ symmetry. However, the valleys in altermagnetic materials are comprised of spin-polarized valleys related by a crystal symmetry instead of the $\mathcal{T}$ symmetry. Ma et al. defined these as crystal symmetry-related valleys ($\mathcal{C}$-paired valleys), which, compared to time-reversal symmetry-related valleys ($\mathcal{T}$-paired valleys), exhibit more intriguing properties in terms of valley polarization~\cite{HMa2021}.
Therefore, applying uniaxial strain to break the crystal symmetry can induce the valley splitting in altermagnetic materials, upon finite doping can further induce a net magnetization ($M$)~\cite{YZhu2024}.

Recent advances in multifunctional 2D piezoelectric materials have spurred extensive experimental and theoretical investigations, positioning them as promising platforms for next-generation multifunctional electronics.~\cite{WWu2014,MDai2019,YGuo2017,LWang2022,ZWang2024,QMa2024,QMa2025}. The piezoelectric effect occurs in semiconductors with non-centrosymmetric crystal structures~\cite{WWu2016,CShi2020}. In 2D altermagnetic materials, such as CrO~\cite{XChen2023}, V$_2$Se$_2$O~\cite{HMa2021}, Fe$_2$Se$_2$O~\cite{YWu2024}, and Cr$_2$Te$_2$O~\cite{QCui2023} have inversion symmetry, leading to missing piezoelectricity. Searching for 2D piezoelectric altermagnetism (PAM) is both significant and challenging. A substantial out-of-plane piezoelectric response is highly desirable for practical applications, as it ensures compatibility with current bottom/top gate technologies. Therefore, altermagnetic materials with a significant out-of-plane piezoelectric response are promising candidates for multifunctional ultrathin piezoelectric devices.

In this work, based on first-principles calculations, we propose an altermagnetic semiconductor: Janus V$_2$AsBrO monolayer. It exhibits an out-of-plane magnetic easy axis and a N\'{e}el temperature ($T_N$) exceeding room temperature. Breaking the crystal symmetry through uniaxial strain induces significant valley splitting, demonstrating a robust piezovalley effect. Furthermore, a strain-tunable piezomagnetic effect emerges under hole doping, generating a finite net $M$.  Additionally, the monolayer exhibits a substantial out-of-plane piezoelectric coefficient (2.19 pm/V), as confirmed by theoretical calculations. The ``multipiezo'' effect combining piezovalley, piezomagnetism, and piezoelectricity establishes the V$_2$AsBrO monolayer an exciting candidate for next-generation multifunctional devices in future nanoelectronics and spintronics.

\section{Methods}

The first-principle calculations are performed using the projector augmented wave (PAW) method within density functional theory (DFT)~\cite{PHohenberg1964,WKohn1965} as implemented in the plane wave code Vienna ab initio Simulation Package (VASP)~\cite{GKresse1995,GhKresse1996,GKressed1999}. The exchange-correlation functional employ the generalized gradient approximation (GGA) formulation of Perdew-Burke-Ernzerhof (PBE)~\cite{JPerdew1996}. The cut-off energy is set to 600 eV, and the convergence criteria for atomic force and energy are set to be 0.001 eV/{\AA} and 10$^{-7}$ eV, respectively. A $\Gamma$-centered Monkhorst-Pack grid of 12 $\times$ 12 $\times$ 1 k-points is used to sample the Brillouin zone (BZ) of the unit cell. A vacuum layer of 20 {\AA} is placed along the $z$ direction to avoid interactions between adjacent layers. To deal with the strong correlation effect of the 3$d$ orbitals of V atom, the GGA + $U$ method ($U$ = 4.0 eV) is employed for the V-3$d$ electron~\cite{XCheng2021,YZhu2024,JY2024}. A 2 $\times$  2 $\times$  1 supercell is adopted to calculate the phonon spectrum by using the PHONOPY code~\cite{ATogo2015} based on the density functional perturbation theory (DFPT)~\cite{XGonze1997}. The $ab$ $inito$ molecular dynamics (AIMD) simulations in the canonical (NVT) ensemble are performed for 3000 fs at 300 K with a Nos\'{e}-Hoover thermostat~\cite{GMartyna1992}.


\section{Results and Discussion}

\begin{figure}[hb!]
\centerline{\includegraphics[width=0.8\textwidth]{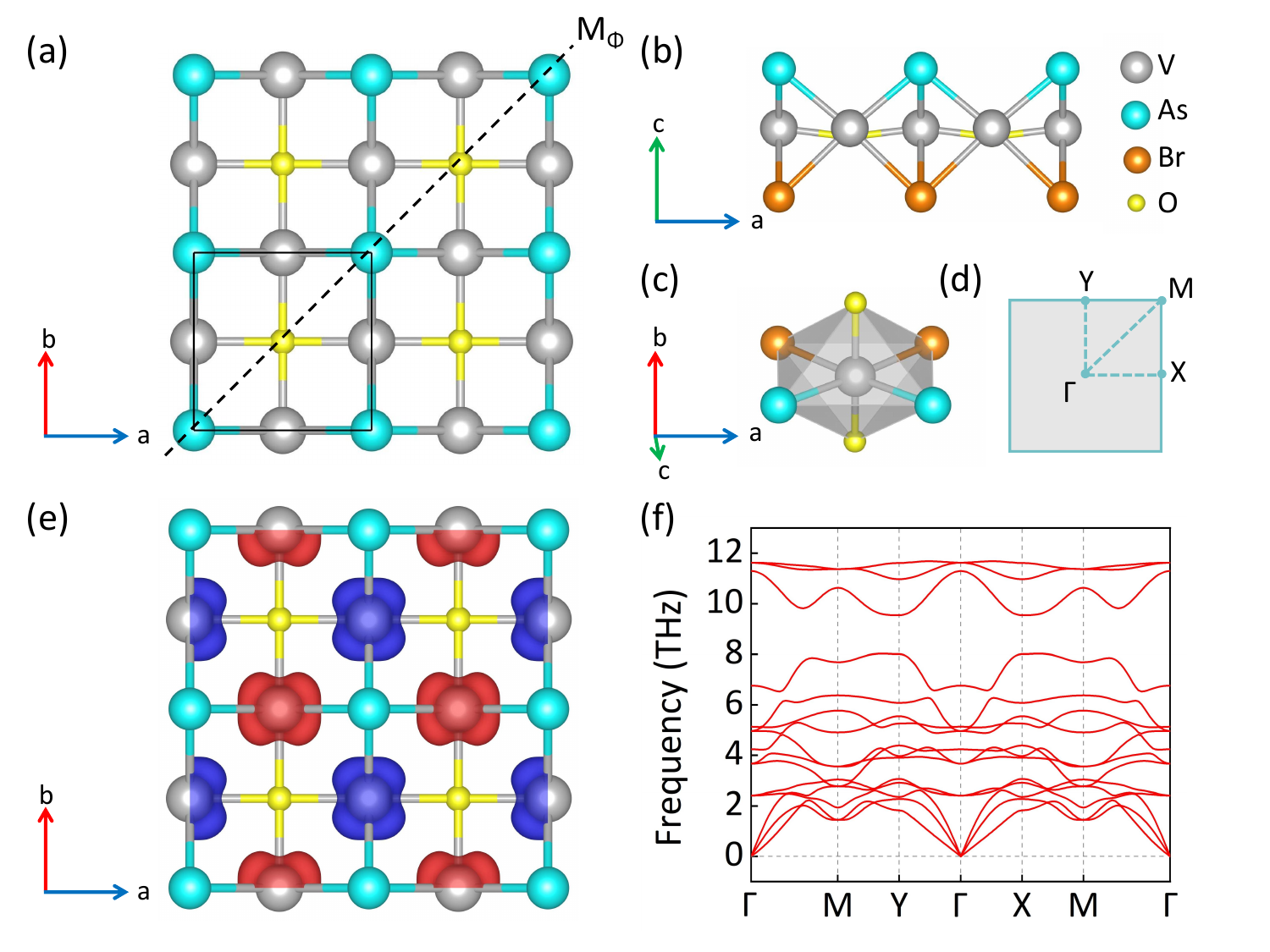}}
\caption{(a) Top and (b) side views of the Janus V$_2$AsBrO monolayer. The unit cell is marked by the solid black line.
M$_\Phi$ denotes a mirror operation performed along a diagonal direction in the $ab$ plane. (c) Schematic diagram of V atoms in an octahedral crystal field for Janus V$_2$AsBrO monolayer. (d) The first Brillouin zone with high-symmetry points. (e) Top views of spin densities of the Janus V$_2$AsBrO monolayer. The spin-up density is shown in red and the spin-down density is shown in blue, respectively.  (f) The phonon dispersion of the Janus V$_2$AsBrO monolayer.
\label{fig:1}}
\end{figure}

Figure~\ref{fig:1}(a-b) illustrates the Janus V$_2$AsBrO monolayer crystal structure, characterized by diagonal mirror symmetry M$_\Phi$. The monolayer adopts a tetragonal lattice (space group $P$4$mm$, No. 99), where a central V-O atomic layer is intercalated between As and Br termination layers. Each V atom is surrounded by two O, two As, and two Br atoms forming a highly distorted octahedral structure [Fig.~\ref{fig:1}(c)]. The Brillouin zone and high symmetry point of Janus V$_2$AsBrO monolayer are depicted in Fig. ~\ref{fig:1}(d). Detailed structural information including lattice constants, bond lengths, and bond angles of Janus V$_2$AsBrO monolayer are summarized in Table S1 of the supplementary material~\cite{SupplMata}. To characterize the bonding nature of the Janus V$_2$AsBrO monolayer, we calculated its electron localization function (ELF), as shown in Fig. S1~\cite{SupplMata}. The ELF distribution reveals pronounced electron density accumulation around O, As, and Br atoms, indicating dominant ionic character in the V-O, V-As, and V-Br bonds. To identify the magnetic ground state, we evaluated four collinear spin configurations: ferromagnetic (FM) and three antiferromagnetic orderings (AFM1, AFM2, AFM3), see Fig. S2~\cite{SupplMata}. Our calculations demonstrate that the AFM1 configuration exhibits the lowest total energy. The energy differences between competing magnetic orders and the AFM1 ground state are systematically tabulated in Table S2~\cite{SupplMata}.

The spin density of the Janus V$_2$AsBrO monolayer exhibits an anisotropic magnetization localized around the V atoms [Fig.~\ref{fig:1}(e)]. Notably, this spin density cannot be manipulated through inversion or translation symmetry, but instead requires a rotational operation, which is a distinguishing feature of altermagnets~\cite{Lmejkalp2022,Lmejkalj2022}. Magnetic anisotropy, a fundamental property of magnetic materials, is essential for achieving long-range ferromagnetic ordering in 2D materials~\cite{XCheng2021,AMa2020}. Anisotropy is typically characterized by the magnetic anisotropy energy (MAE), which quantifies the energy variation with respect to the magnetization direction. The calculated MAE (MAE = E$_{[100]}$ - E$_{[001]}$) is 110 $\mu$eV per unit cell, indicating a perpendicular magnetic anisotropy (PMA) character in the Janus V$_2$AsBrO monolayer. The strong PMA character is further demonstrated by the MAE distribution over the entire space, as shown in Fig. S3(a)~\cite{SupplMata}. Specifically, the MAE reaches its maximum value in the $xy$-plane and gradually decreases to zero as the magnetic axis rotates toward the $z$-direction. In addition, we calculated the $T_N$ of the Janus V$_2$AsBrO monolayer to be 960 K [Fig. S3(b)~\cite{SupplMata}] using the Monte Carlo (MC) method based on the Heisenberg model. (More details can be found in the supplementary material.) The value of $T_N$ is significantly higher than that of other altermagnets, such as Fe$_2$Se$_2$O (319 K)~\cite{YWu2024},  Mn$_2$Se$_2$O (467 K)~\cite{FHuang2024}, and V$_2$Te$_2$O (740 K)~\cite{QCuiui2023}.

To determine the stability of Janus V$_2$AsBrO monolayer, we performed first-principles calculations of phonon dispersion, AIMD simulations, and elastic tensor analysis. The phonon dispersion in Fig.~\ref{fig:1}(f) exhibits no imaginary frequencies throughout the entire Brillouin zone, confirming its dynamic stability. Furthermore, AIMD simulation at 300 K reveal minimal energy fluctuations ($\triangle$$E$ $<$ [0.15] eV/atom) and preserved structural integrity over [3] ps trajectories (see Fig. S4~\cite{SupplMata}), demonstrating robust thermal stability. The elastic constants ($C_{11}$ = 94.08 N/m, $C_{12}$ = 20.57 N/m, and $C_{66}$ = 34.65 N/m) of Janus V$_2$AsBrO monolayer satisfy the Born-Huang criteria for mechanical stability~\cite{RAndrew2012}: $C_{11} > 0$, $C_{66} > 0$, $C_{11} > \left| C_{12} \right|$, confirming its mechanical stability. Besides stability, it is important to consider material strength and anisotropy for practical applications. Therefore, we calculated the angular-dependent elastic properties: Young's modulus $Y_{2D}$($\theta$) and Poisson's ratio $\nu_{2D}$($\theta$) (Fig. S5~\cite{SupplMata}). The max (min) Young's modulus of Janus V$_2$AsBrO monolayer is 89.58 N/m (86.37 N/m). It is noteworthy that the Young's modulus of V$_2$AsBrO is lower than that of MoS$_2$ (124.5 N/m)~\cite{DPeeter2014}, suggesting it can be easily tuned by strain. This makes the Janus V$_2$AsBrO monolayer highly suitable for novel flexible electronics and piezotronics applications. Besides, the max (min) Poisson's ratio of 0.24 (0.20) is almost equal to that in graphene (0.19)~\cite{APolitano2012}.

\begin{figure}[t!hp]
\centerline{\includegraphics[width=1.05\textwidth]{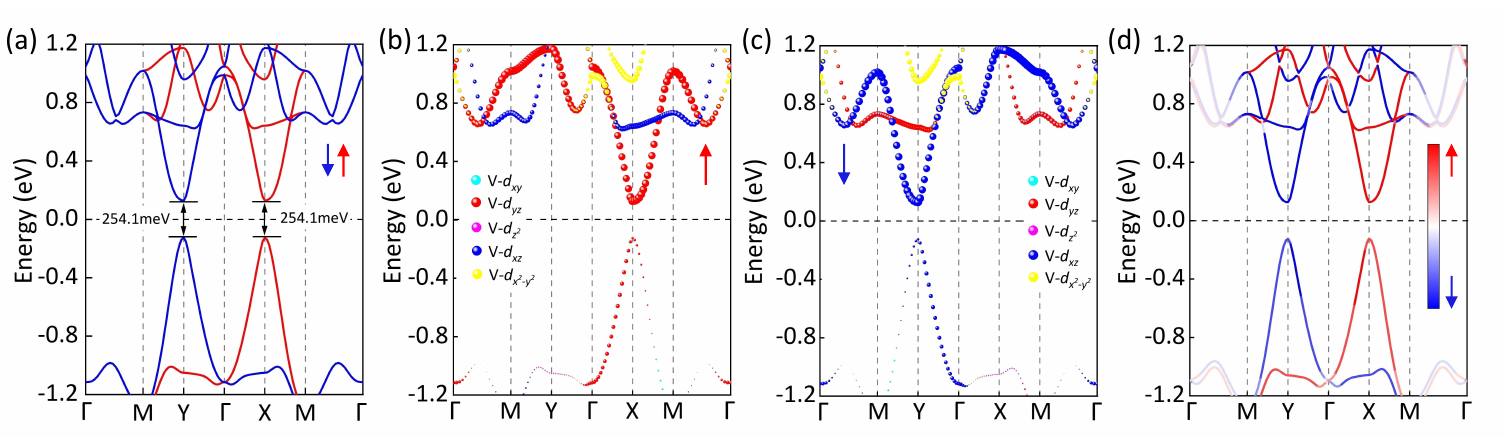}}
\caption{(a) Spin-polarized electronic band structure of the Janus V$_2$AsBrO monolayer in the AFM ground state, calculated without SOC. (b, c) The V $d$-orbital-projected band structures for spin-up and spin-down channels (SOC excluded). (d) SOC-incorporated spin-resolved band structure, demonstrating preserved valley degeneracy.
\label{fig:2}}
\end{figure}

We systematically investigate the electronic properties of the Janus V$_2$AsBrO monolayer. As shown in Fig.~\ref{fig:2}(a), the spin-resolved band structure without SOC reveals a semiconductor character with a direct bandgap of 254.1 meV. The energy band structure shows the splitting of opposite spins in reciprocal space, confirming the characteristic feature of altermagnets. Mirror-symmetry-related valley degeneracy is identified at the $Y$ and $X$ high-symmetry points in both conduction and valence bands. Figure~\ref{fig:2}(b-c) presents the projected band structures for the V atoms demonstrating  that the $Y$/$X$ valleys in spin-up channels are dominated by $d_{yz}$ orbitals, while $d_{xz}$ orbitals primarily contribute to spin-down channels.

Incorporating SOC effects [Fig.~\ref{fig:2}(d)],  the valley degeneracy remains preserved, indicating that valley polarization in this altermagnet originates from spin splitting rather than SOC. To further verify the SOC effect, we calculated the band structures with SOC under [100] and [010] magnetization directions (Fig. S6~\cite{SupplMata}). While minor valley polarization ($\triangle$$E$$_{v}$ $<$ [5] meV) emerges in these configurations, the negligible magnitude confirms the weak SOC-valley coupling. Consequently, SOC is excluded in subsequent calculations to isolate intrinsic altermagnetic effects.

\begin{figure}[t!hp]
\centerline{\includegraphics[width=0.81\textwidth]{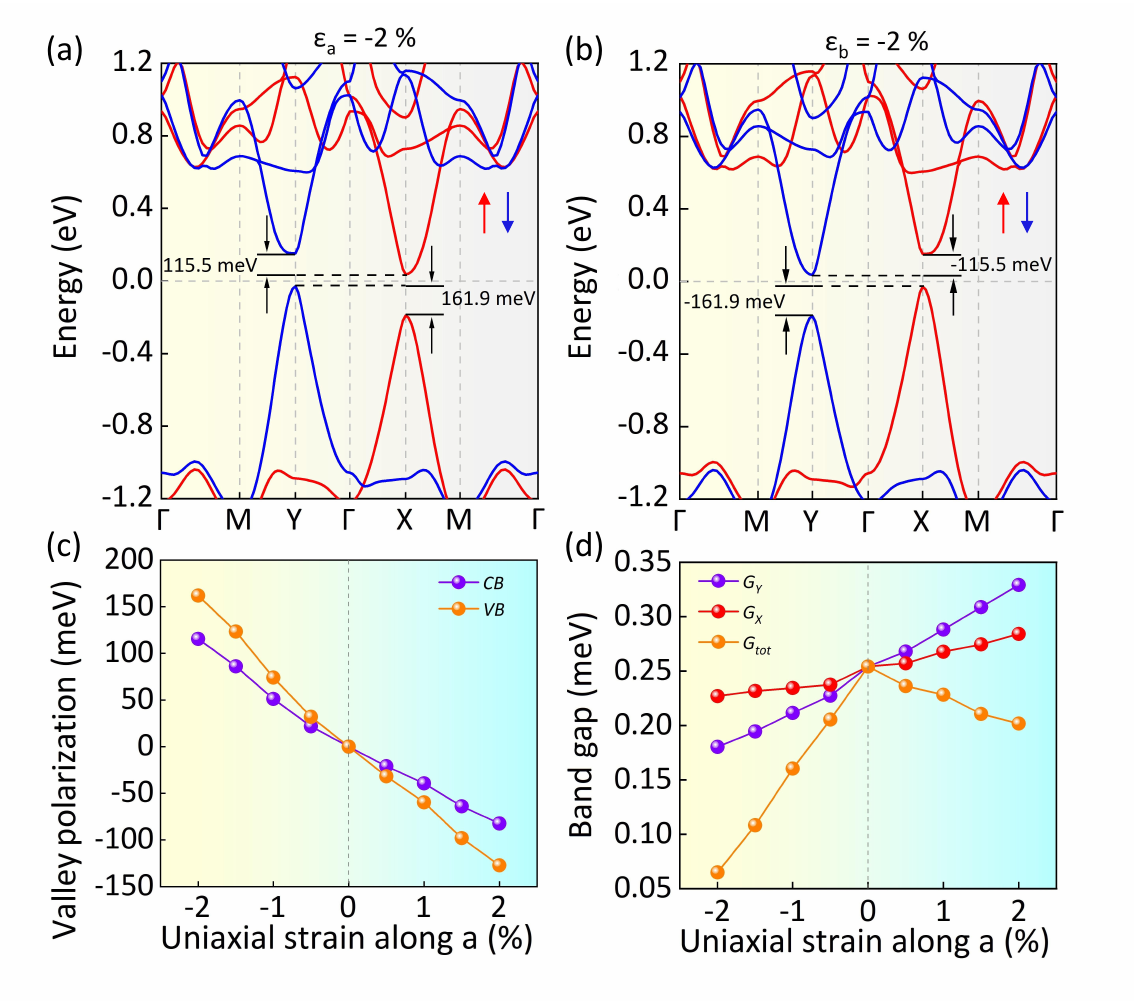}}
\caption{Band structures of the Janus V$_2$AsBrO monolayer under $-2\%$ uniaxial compressive strain along (a) $a$ direction and (b) $b$ direction. (c) Strain-induced valley polarization in the conduction band ($CB$) and valence band ($VB$) under $a$ direction uniaxial strain. (d) Evolution of valley gaps at $Y$ and $X$ [$G_{Y}$ and $G_{X}$], as well as the total bandgap [$G_{tot}$] as function of  $a$ direction strain.
\label{fig:3}}
\end{figure}

The piezovalley effect manifests as a strain-mediated valley polarization mechanism, arising from the coupling between lattice deformation and spin-valley degrees of freedom in noncentrosymmetric crystals. Due to the M$_\Phi$ symmetry of the Janus V$_2$AsBrO monolayer, the two valleys at $Y$ and $X$ are degenerate. In the V$_2$AsBrO monolayer, the M$_\Phi$ symmetry enforces energy degeneracy between the $Y$ and $X$ valleys. Therefore, one can use in-plane uniaxial strain along the $a$ or $b$ directions to break the crystal symmetry and induce substantial valley polarization. As shown in Figs.~\ref{fig:3}(a) and ~\ref{fig:3}(b), the V$_2$AsBrO monolayer at $-2\%$ uniaxial strains along the $a$ and $b$ directions makes the energy of $Y$ valley lower/higher than that of $X$ valley. The value of valley polarization is 161.9 (155.5) meV at the valence band (conduction band) under 2\% uniaxial strain along $a$ direction [Fig.~\ref{fig:3}(a)], which is significantly larger than the reported values in 2D conventional ferrovalley materials requiring $\mathcal{T}$ symmetry~\cite{CZhang2019,PJiang2021,PZhao2022,BHuang2023}. The Janus V$_2$AsBrO monolayer exhibits valley polarization of $-$161.9 ($-$155.5) meV at the valence band (conduction band) under $-2\%$ uniaxial strain along the $b$ direction [Fig.~\ref{fig:3}(b)], indicating that the sign of valley polarization can be reversed by applying the uniaxial strain. As shown in Fig.~\ref{fig:3}(c), the valley polarization in the conduction band ($CB$) and valence band ($VB$) changes increase linearly under 0 to $-2\%$ and 0 to 2.0\% uniaxial strains along the $a$ direction, which can expands the scope of the study of valley materials. The energy difference ($\Delta$$E$) and band structures of the V$_2$AsBrO monolayer under $-2\%$ to 2.0\% uniaxial strain, which indicate that the Janus V$_2$AsBrO monolayer retains its altermagnetic properties, are presented in Fig. S7 $\sim$ S9~\cite{SupplMata}. In addition, Fig.~\ref{fig:3}(d) shows the variation in the overall band gap [$G_{tot}$] and the gaps of the $Y$ and $X$ valleys [$G_{Y}$ and $G_{X}$]. It is clear that uniaxial strain promotes the separation of the $Y$ and $X$ valleys in both the conduction and valence bands from other bands, facilitating the experimental manipulation of these valleys. Furthermore, biaxial strain is also considered to adjust and control the electronic structure. The calculations indicate that the Janus V$_2$AsBrO monolayer retains its antiferromagnetic property, and the band gap gradually increases with the application of biaxial strain (see Figs. S10 and S11~\cite{SupplMata}).

\begin{figure}[t!hp]
\centerline{\includegraphics[width=0.8\textwidth]{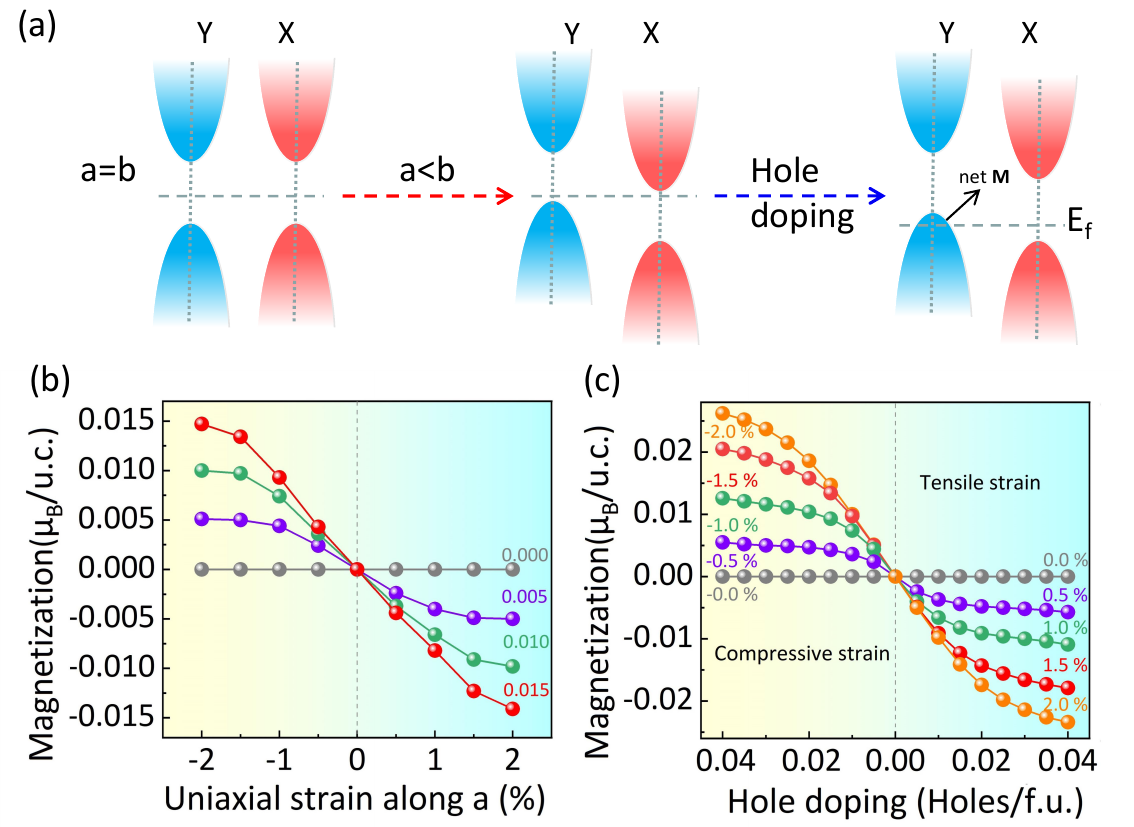}}
\caption{ (a) Schematic of the net magnetization ($M$) induced by hole doping in the Janus V$_2$AsBrO monolayer under certain uniaxial strain. (b) and (c) The corresponding net magnetization per unit cell for different concentrations of hole doping.
\label{fig:4}}
\end{figure}

The physical mechanism underlying the piezomagnetic effect in the V$_2$AsBrO monolayer is illustrated in Fig.~\ref{fig:4}(a). Upon the manifestation of valley polarization in the band structure due to uniaxial strain, a net $M$ emerges by precisely aligning the Fermi level to intersect a single valley through hole doping. The $M$ is elucidated by the formula: $M = \int_{-\infty}^{E_F^{(n)}} \left[ \rho^\uparrow(\varepsilon) - \rho^\downarrow(\varepsilon) \right] dE$, where $E_F$ is the fermi level after doping, $n$ is the doping density, and $\rho^\uparrow (\rho^\downarrow)$ is the spin-up (down) part of the density of states. Subsequently, an evaluation was conducted to determine the resultant magnetic moment $M$ for distinct strains and varying hole doping levels. As shown in Fig. \ref{fig:4}(b), under a certain hole doping concentration, the net $M$ increases consistently with increasing uniaxial strain and eventually saturates. Notably, tensile and compressive strains induce opposite magnetic orientation. Analogously, as shown in Fig. \ref{fig:4}(c), for a certain uniaxial strain, the net $M$ exhibits a gradual rise with increasing hole doping concentration, reaching a maximum saturation point with heightened doping levels. The band structures of V$_2$AsBrO monolayer subjected to uniaxial strain along the $a$ direction, doped with 0.03 hole/f.u. are presented in Fig. S12~\cite{SupplMata}, thereby demonstrating the variation of net $M$ polarization.

The piezoelectric effect is an intrinsic electromechanical coupling phenomenon in non-centrosymmetric materials. The application of strain or stress causes a charge redistribution, thereby leading to the formation of electric dipole moments and the production of electricity. Owing to the disruption of horizontal mirror symmetry, the Janus V$_2$AsBrO monolayer exhibits an out-of-plane piezoelectricity. The 2D manifestation of the piezoelectric effect is described by the piezoelectric stress coefficients ($e_{ij}$) and the piezoelectric strain coefficients ($d_{ij}$). The relaxed piezoelectric tensors ($e_{ij}$ and $d_{ij}$) are obtained as the sum of electronic and ionic contributions:
\begin{eqnarray}
{\emph{e$_{ij}$} = \frac{\partial{P_{i}}}{\partial{\varepsilon_{j}}} = e_{ij}^{elc} + e_{ij}^{ion}},
\end{eqnarray}
\begin{eqnarray}
{\emph{d$_{ij}$} = \frac{\partial{P_{i}}}{\partial{\sigma_{j}}} = d_{ij}^{elc} + d_{ij}^{ion}},
\end{eqnarray}
where the \emph{P$_{i}$}, ${\varepsilon_j}$, and $\sigma_{j}$ represent the piezoelectric polarizations, strains, and stresses, respectively. Using Voigt notation, the piezoelectric strain and piezoelectric stress tensors with C$_{4v}$ point group can be expressed as:
\begin{eqnarray}
	\begin{pmatrix}
	 0 & 0 & 0 \\
	0 & 0 & 0 \\
	e_{31} & e_{31} & 0
	 \end{pmatrix},
\end{eqnarray}
\begin{eqnarray}
	\begin{pmatrix}
	 0 & 0 & 0 \\
	0 & 0 & 0 \\
	d_{31} & d_{31} & 0
	 \end{pmatrix}.
\end{eqnarray}

The piezoelectric strain coefficients \emph{d$_{ik}$} can be derived by piezoelectric stress coefficients \emph{e$_{ik}$} and elastic stiffness coefficients \emph{C$_{ik}$}:
\begin{eqnarray}
{\emph{e$_{ik}$} = d_{ij}C_{jk} },
\end{eqnarray}
therefore, the out-of-plane piezoelectric coefficients \emph{d$_{31}$} can be expressed as:
\begin{eqnarray}
{\emph{d$_{31}$} = \frac{{e_{31}}}{{C_{11}} + {C_{12}}} }.
\end{eqnarray}

The \emph{e$_{31}$} of Janus V$_2$AsBrO monolayer can be directly calculated by using DFPT. The calculated \emph{e$_{31}$} (2.52 $\times$ $10^{-10}$ C/m) mainly arises from the contribution of electrons, and the signs for the contributions of electrons (2.84 $\times$ $10^{-10}$ C/m) and ions (-0.32 $\times$ $10^{-10}$ C/m) in Janus V$_2$AsBrO monolayer is opposite. According to equations (6), the calculated \emph{d$_{31}$} is 2.19 pm/V, which is significantly larger than that of the altermagnetic semiconductor Janus Cr$_2$SO monolayer ($\left|\emph{d$_{31}$} \right|$ = 0.97 pm/V)~\cite{SGuo2023}. Notably, the \emph{d$_{31}$} of Janus V$_2$AsBrO monolayer is over nine times greater than that of Janus V$_2$SeTeO monolayer ($\sim$ 0.24 pm/V)~\cite{YZhu2024}. The large \emph{d$_{31}$} provides the potential to tune the spin-split bands of the altermagnetic V$_2$AsBrO monolayer via the piezoelectric effect, offering broad prospects for the development and design of novel piezoelectric devices. Furthermore, as shown in Fig. S13~\cite{SupplMata}, we calculated the planar averaged electrostatic potential energy variation along the $z$ direction and the intrinsic polar field of the V$_2$AsBrO monolayer to reveal the origin of its piezoelectricity. The difference in electrostatic potential energy between the sides is 1.08 eV, which can be interpreted as a surface-dependent work function, indicating charge redistribution within the V$_2$AsBrO monolayer, as well as the existence of an out-of-plane dipole moment. The strength of the intrinsic polar field was deduced from the slope of the curve, approximately 0.37 eV/{\AA}. The out-of-plane dipole and intrinsic polar field indicate internal polarization of the material, further suggesting the potential piezoelectricity of the Janus V$_2$AsBrO monolayer.

\section{Conclusion}
In conclusion, we predict a 2D altermagnetic semiconductor Janus V$_2$AsBrO monolayer with a band gap of 254.1 meV based on first-principles calculations.
The material exhibits excellent dynamic, thermal, and mechanical stability, as confirmed by phonon dispersion spectra, ab initio molecular dynamics simulations, and elastic constant analysis. Remarkably, the monolayer demonstrates a strong PMA character with the MAE of 110 $\mu$eV per unit cell, and an ultrahigh T$_N$ of 960 K derived from MC simulations, far exceeding room-temperature operational requirements. Uniaxial strain engineering in the V$_2$AsBrO monolayer induces sizeable valley splitting, revealing the piezovalley effect. By further introducing hole doping under strain, a net magnetization can be generated, demonstrating a robust piezomagnetic effect. Additionally, the breaking of horizontal mirror symmetry results in a large out-of-plane piezoelectric coefficient \emph{d$_{31}$} = 2.19 pm/V, which significantly exceeds those of other predicted 2D piezoelectric altermagnets.
These multifunctional properties-altermagnetism, piezovalley polarization, and piezoelectricity are synergistically integrated into a single material, enabling unprecedented control over spin, valley, and mechanical degrees of freedom. Our findings highlight the Janus V$_2$AsBrO monolayer as a promising candidate for next-generation multifunctional devices in spintronics, valleytronics, and piezoelectric applications.
\section{Acknowledgments}

This work was supported by the Natural Science Foundation of China under Grants (No. 22372142), the Innovation Capability Improvement Project of Hebei province (No. 22567605H), the Natural Science Foundation of Hebei Province of China (No. B2021203030), the Science and Technology Project of Hebei Education Department (No. JZX2023020). The numerical calculations in
this paper have been done on the supercomputing system in the High Performance Computing Center of Yanshan University.



\end{document}